\documentclass[
 reprint,
 aps,
 superscriptaddress
]{revtex4-1}

\usepackage{graphicx}

\begin{document}

\title{Anisotropic micro-cloths fabricated from DNA-stabilized carbon nanotubes:\\one-stop manufacturing with electrode needles
}
\author{Hiroshi Frusawa}
\email{frusawa.hiroshi@kochi-tech.ac.jp}
\author{Gen Yshii}
\affiliation{Institute for Nanotechnology, Kochi University of Technology, Tosa-Yamada, Kochi 782-8502, Japan.}


\begin{abstract}
Among a variety of solution-based approaches to fabricate anisotropic films of aligned carbon nanotubes (CNTs), we focus on the dielectrophoretic assembly method using AC electric fields in DNA-stabilized CNT suspensions. We demonstrate that a one-stop manufacturing system using electrode needles can draw anisotropic DNA-CNT hybrid films of 10-100 $\mu$m in size (i.e., free-standing DNA-CNT micro-cloths) from the remaining suspension into the atmosphere while maintaining structural order.
It has been found that a maximal degree of polarization (ca. 40 \%) can be achieved by micro-cloths fabricated from a variety of DNA-CNT mixtures.
Our results suggest that the one-stop method can impart biocompatibility to the downsized CNT films and that the DNA-stabilized CNT micro-cloths directly connected to an electrode could be useful for biofuel cells in terms of electron transfer and/or enzymatic activity.\\

\end{abstract}

\maketitle

\section*{Background}

Over the last decade, a variety of solution-based methods have been developed to fabricate carbon nanotube (CNT) films [1-23] due to their potential application as electronic (e.g., thin film transistors) [1,3,4,17-19] or electrochemical devices (e.g., biofuel cells [2-4,24-28] and biosensors [24,29-32]).
In biofuel cells, efficient electron transfer is required between the enzymatic active site and the electrode surface.
Metallic CNTs, including DNA-wrapped CNTs [27],  are ideal conducting nanowires that would facilitate such electron transfer.
The difficulty, however, lies in the incorporation of CNT films into such devices, because the anisotropic CNT films of submicron to micron scales need to be manipulated for optimizing the device layout.

Solution-based approaches used to prepare CNT films mainly stem from colloidal assembly methods [1-28,33], and are the most popular and versatile techniques for fabricating composite films [5,20-23].
The main advantage of the colloidal method is that it can yield various films directly at room temperature in a manner compatible with thermal, piezoelectric, or electro-hydrodynamic jet printing patterning techniques [1-4].
However, this approach also has several disadvantages as it depends on the wettability of the substrates limiting the range of viable substrates for various applications.
Additionally, the force involved in the transfer process easily disrupts the original network, causing limitations in terms of the film quality and the production efficiency.
Colloidal approaches thus need to be refined to incorporate the aligned arrays of CNTs into various devices. 

To this end, we used the dielectrophoretic assembly method [12-19] using AC electric fields in DNA-stabilized CNT suspensions [11,23,25-27,33-36] to produce aligned arrays of CNTs.
The dielectrophoretic method provides a notable advantage over other techniques because this electronic method is able to selectively collect metallic CNTs and separate them from semiconducting CNTs [15,16].
Earlier studies have demonstrated that AC fields applied via on-chip electrodes precisely control the orientation of CNT arrays [12-19].
However, the on-chip assembly systems developed so far have used electrode configurations fixed to the substrate, making electrode removal from the CNT assembly difficult without use of a binder [37].

In this study, we overcome the difficulties in fixed electrode configuration using plug-in electrode needles inserted into a pre-assembly suspension above the substrate.
Using this method, we prepared free standing anisotropic DNA-CNT hybrid films shaped into rectangular micro-cloths of 10-to 100-$\mu$m in size using electrode needles with 0.5-$\mu$m tip diameters.
Our aim is to demonstrate that the use of plug-in electrodes solves several of the main issues found in wet fabrication methods.
Furthermore, our method allows hybrid CNT film assembly to be drawn from a suspension into a gaseous atmosphere in one-stop manufacturing process while maintaining structural order.

\section*{Methods}

\subsection*{Materials}

Denatured DNA has been proven to be efficient to stabilize CNT densities lower than 1 wt.\% in water [11,23,25-27,33-36].
We first prepared aqueous solutions of single-walled CNTs (purity, $>95 \%$; Cheap Tubes) using salmon DNA (WAKO) as follows:
10 mg of raw CNT soot was suspended in 5 ml of a 0.12 wt.\% density solution of salmon DNA (Wako) denatured by 10-min heating at $90\,^\circ\mathrm{C}$.
The DNA-CNT mixture was sonicated for 60 min at a 20 kHz frequency at 50 W in a water-ice bath using a homogenizer (Branson, Sonifier 250).
The sonicated suspension was centrifuged at 650g for 10 min, and the carefully decanted supernatant was used for the dielectrophoretic assembly.
The CNT concentration in the supernatant was estimated from measuring its dry weight.
We further added DNA solutions at various densities to the supernatants of the DNA-stabilized CNTs.

In performing the dielectrophoretic assembly experiments, we used diluted DNA-CNT mixtures where the CNT density, $C_{\mathrm{CNT}}$, was adjusted at $C_{\mathrm{CNT}}=0.1$ wt.\%.
The concentrations of added DNA, $C_{\mathrm{DNA}}$, adopted in the assembly ranged from 0.06 wt.\% to 0.2 wt.\%.
Accordingly, we investigated the effect of density ratio, $\alpha=C_{\mathrm{CNT}}/C_{\mathrm{DNA}} $, in the range of $0.6\leq\alpha\leq 2$ while maintaining the CNT density at $C_{\mathrm{CNT}}=0.1$ wt.\%.
It is to be noted that the formation of a nematic phase has been reported at $\alpha=1$ in more concentrated suspensions ($C_{\mathrm{CNT}}>1$ wt.\%) [11,33-36].

\subsection*{Experimental Setup}

A pair of tungsten needles with a tip diameter of 0.5 $\mu$m were independently controlled by two sets of patch clamp micromanipulators (Narisige, NMN-21).
The electrode needles were inserted into a 10-$\mu$L aliquot of CNT suspensions mounted on an inverted optical microscope (Olympus, IX71) and were arranged in a parallel configuration.
An external sinusoidal wave electric field was applied between the electrode needles using an arbitrary waveform generator (Agilent, 33220A) along with a current amplifier (FLC Electronics, F30PV).
We adjusted two parameters of the AC fields: applied electric field strength and wave frequency.
Electric fields applied were in the range of 25-200 kV/m while frequencies applied were between 1 kHz and 20 MHz.
We obtained a time series of assembling images via CCD  camera (Q-Imaging, Retiga Exi), and performed video analysis using image analysis software (ImageJ).

Observations of the prepared CNT films were also made using a digital microscope (Omron, VC7700), and a field emission scanning electron microscope (FE-SEM; JEOL, JSM-7300F) operated at 15 kV.
We also covered the substrate glass with black polymer film. Within the black polymer film, a 70-$\mu$m pinhole was created.
Through the pinhole placed on a microscope stage, the dried film was observed with its edge attached to a needle.
The degree of alignment was quantified using a polarizer (Olympus), a spectrometer (Ocean Optics, USB2000+), and a halogen light source with its peak intensity at approximately 600 nm. 

\section*{Results and discussion}

\subsection*{Dispersion properties of DNA-CNT mixtures}

The DNA-stabilized CNTs [11,23,25-27,33-36]  disperse and stabilize at a considerable range of concentrations.
Therefore, an isotropic-to-nematic phase transition has been found to occur in concentrated DNA-wrapped CNT suspensions particularly at the optimal ratio of $\alpha=1$ [11,33-36].
This is similar to the transition observed in lyotropic rigid-rod polymers [38].
In our experiments, six different DNA-CNT mixtures were prepared by changing the density ratio $\alpha$ while maintaining $C_{\mathrm{CNT}}=0.1$ wt.\%.
This value is much lower than the density region ($C_{\mathrm{CNT}}>1$ wt.\%) where the nematic phase had been found to emerge [11,33-36];
accordingly, optical microscopy was unable to detect the emergence of a nematic order in our dilute pre-assembly suspensions.

\begin{figure}[hbtp]
\begin{center}
	\includegraphics[
	width=6.8cm
	]{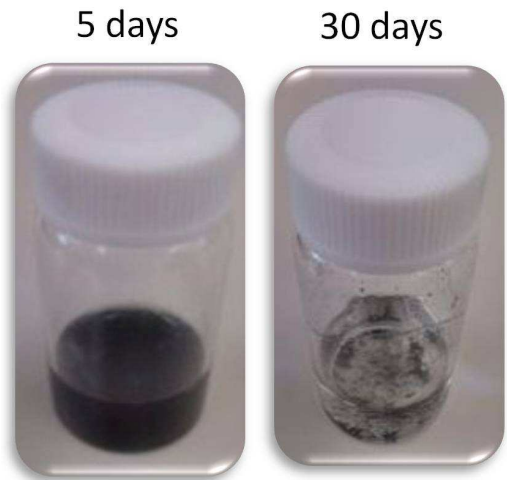}
	\end{center}
	\caption{{\bf Photographs of DNA-CNT mixtures.} Uniform and sedimentary dispersions were observed after 5-and 30-day incubation periods, respectively, at $C_{\mathrm{CNT}}=0.1$ wt.\% and $\alpha=2$.
}
\end{figure}

All of the relatively dilute dispersions appeared uniform and optical micrographs displayed few aggregates (see also Movie 1 in Additional File 2) for several days.
Based on these observations, we performed CNT micro-cloth fabrications using dispersions incubated at $20\,^\circ\mathrm{C}$ for 24 hours.
Thereafter, the DNA-CNT mixtures were left at $20\,^\circ\mathrm{C}$ for 30 days.
After this period, differences were observed.
First, while long-lasting uniformity was validated for $\alpha<1$, we saw an apparent density gradient at $\alpha =1$.
Second, deposited droplets of aggregates were found at $\alpha= 2$ (see Figure 1).
This value $\alpha=1$ agrees with the isotropic-nematic transition line previously reported for dense DNA-CNT mixtures [11,33-36].
The correspondence can be attributed to the enhancement of depletion attraction [38] between CNTs with increasing $\alpha$, which would ultimately leads to sedimentation.
Indeed, the osmotic compression has been observed due to added polymers inducing separation of a CNT-rich anisotropic phase in equilibrium with an isotropic one at $\alpha=1$ [33,35].

\subsection*{One-stop manufacturing of CNT micro-cloths}

The one-stop manufacturing process consists of three consecutive steps:
electric gathering and alignment of DNA-stabilized CNT arrays in suspensions of various DNA-CNT mixtures (Step 1), solvent evaporation following rinsing of the remaining dispersions (Step 2), and separation from the electrodes  (Step 3).
We describe details of this fabrication procedure using the one-stop preparation at $\alpha=1$ (see also Movie1-2 in Additional File 2) as an example.

Figure 2a shows a schematic diagram of CNTs gathering between a parallel pair of inserted electrode needles.
The electrode pair creates an electrically focused rectangle by applying a spatially inhomogeneous AC field.
We set the rectangular length and width at 100 $\mu$m and 300 $\mu$m, respectively (Step 1).
After CNTs were collected in the electric well of the rectangular box, extra CNT dispersoids surrounding the electrodes were rinsed off using deionized water.
Subsequently, natural drying was allowed to proceed.
During this drying process, the AC field was maintained to reduce the alignment loss due to evaporation induced solvent flow (Step 2).
The electrodes were independently controlled by two micromanipulators.
This facilitated not only needle insertion into suspensions while lifting them off the substrate, but also the gentle separation of the dry assembly from the electrodes one by one (Step 3).

The suspended configuration of the electrode needles was useful to avoid the additional deposition of aggregates onto the genuine assembly during drying.
Indeed, even without further rinses, the presence of extra sediments on the surface was rarely observed.
Figures 2b,c display digital microscope images of dried CNT micro-cloths fabricated at $E=25$ kV/m and $f=20$ MHz.
Figure 2b shows a dried CNT micro-cloth with one side attached to an electrode, and Figure 2c presents an isolated sheet on a new cover slip.
From Figure 2c, the dry film can be determined to be approximately 100 $\mu$m in width and 200 $\mu$m in length.
This is somewhat smaller in length than rectangular scale established by the electrode pair.
The optical micrographs in Figures 2b,c reveal that the microcloth edges on both images are bow-shaped due to vaporization flow, which can be clearly seen from Movie 1-2 in Additional File 2.
This vaporization accounts for the decrease in length of the rectangular micro-cloths. 

\begin{figure}[hbtp]
\begin{center}
	\includegraphics[
	width=8.5cm
	]{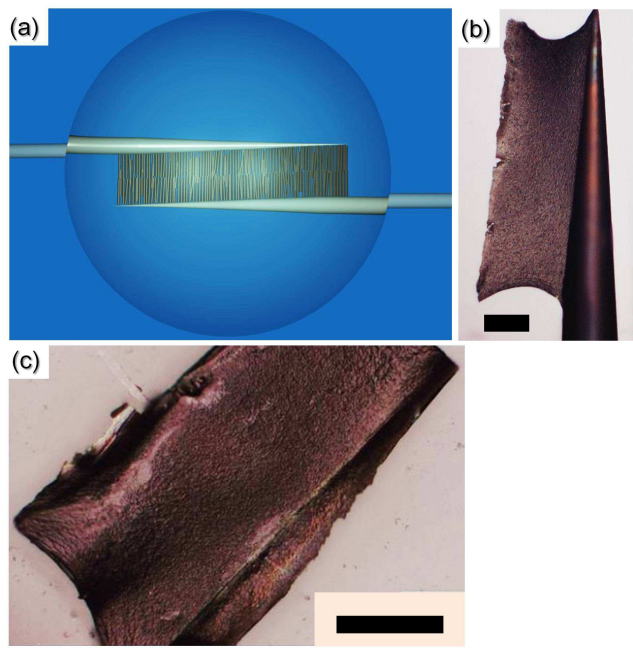}
	\end{center}
	\caption{{\bf Fabrication of CNT micro-cloths using electrode needles.} ({\bf a}) A schematic cofiguration of the electrode needles. These electrode needles are inserted into a suspension of DNA-CNT mixture, collecting aligned CNTs between them. ({\bf b, c}) Digital microscope images of a dried CNT film with its edge attached to one of the electrode needles ({\bf b}), and another sheet placed on a substrate after separation from the electrode pair ({\bf c}). Scale bars: 50 $\mu$m.
}
\end{figure}

\subsection*{Anisotropic properties of CNT micro-cloths}

We used the FE-SEM  to observe aligned CNT arrays inside an assembly film placed on a Cu grid substrate in the same manner as that of Figure 2c.
The left image in Figure 3 shows an FE-SEM image of the surface, whereas a superficial rent is magnified in the right micrograph of Figure 3.
These FE-SEM images reveal that the CNTs were aligned in parallel to the direction of the applied electric fields.
From the adjustment of the SEM focus when observing dry CNT films, we found that the mean CNT film thickness was 5-10 $\mu$m [39]. 

\begin{figure}[hbtp]
\begin{center}
	\includegraphics[
	width=8.5cm
	]{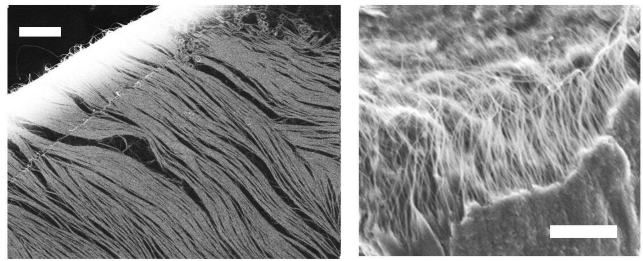}
	\end{center}
	\caption{{\bf FE-SEM images of DNA-stabilized CNT arrays.}
The aligned CNTs constitute electrically fabricated micro-cloths. Scale bars: 2 $\mu$m.
}
\end{figure}

To verify this anisotropy over a larger scale, the alignment degree was quantified using a polarized optical microscope. We performed in situ measurement of the degree of (DOP) for a CNT micro-cloth with one edge to an electrode (see Figure 2b and Movie 1-2 in Additional File 2). The manipulability of the electrode needle allows the film position to be adjusted to overlap a pinhole, as shown in Movie 3 in Additional File 2. Upon observing the anisotropy through the pinhole, the transmitted light intensity was modulated by the rotation of a polarizing plate (see Movie 3 in Additional File 2 for an example).

We measured the optical spectra of the transmitted light at the CCD plane using the aforementioned spectrometer so that we could quantify the maximum and minimum light intensities at the wavelength of 600 nm, $I_{\mathrm{max}}$ and $I_{\mathrm{min}}$, respectively, to obtain the DOP: $\rho = (I_{\mathrm{max}}-I_{\mathrm{min}})/(I_{\mathrm{max}}+I_{\mathrm{min}})$.
A typical set of optical spectra with a given $I_{\mathrm{max}}$ and $I_{\mathrm{min}}$ is shown in Figure 4a where the corresponding micrograph of the brightest micro-cloth was observed through the pinhole.
From these data, the DOP of a CNT micro-cloth fabricated from DNA-CNT mixture at $\alpha=1$ was found to be $\rho\approx42 \%$.

\begin{figure}[hbtp]
\begin{center}
	\includegraphics[
	width=7.5cm
	]{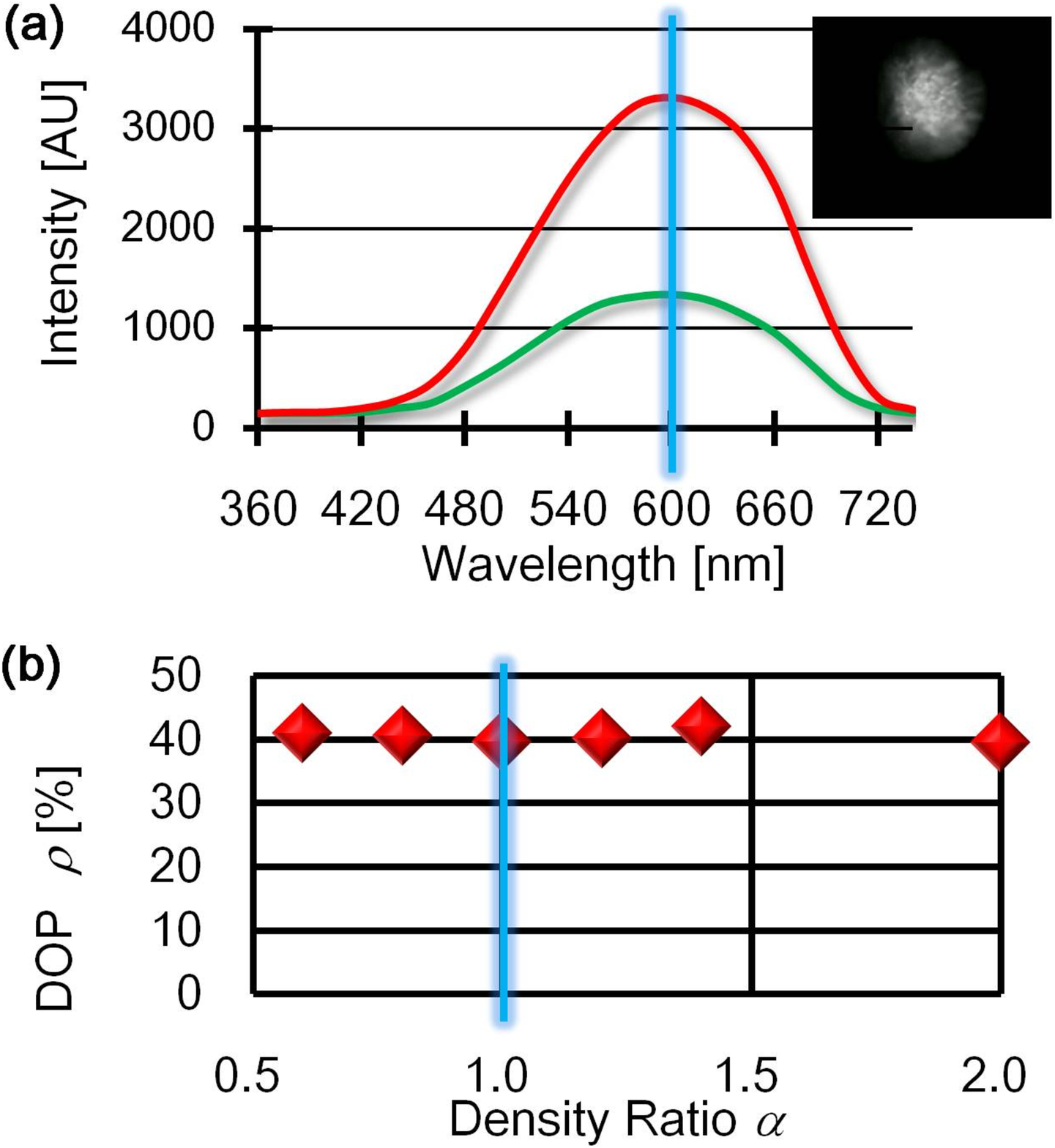}
	\end{center}
	\caption{{\bf Evaluating polarization degrees of CNT micro-cloths.}
({\bf a}) Optical spectra of transmitted light intensities provide $I_{\mathrm{max}}$ and $I_{\mathrm{min}}$ that are evaluated from red and green lines, respectively, at the wavelength of 600 nm marked by the blue line. The spectrum of red line reflects the micrograph of a manipulated CNT micro-cloth observed through a pinhole (see also Movie 3 in Additional File 2).
({\bf b}) DNA concentration dependencies on the DOP ($\rho$) at $E = 25$ kV/m and $f = 20$ MHz. The shaded blue line at $\alpha=1$ marks that DNA-CNT mixtures form depletion-induced bundles above $\alpha=1$ in correspondence with the emergence of nematic order previously reported. Error bars lie within symbols.
}
\end{figure}

We then investigated the extent to which the DOP is correlated with local directional order in the pre-assembly suspension by changing ƒ¿ from 0.6 to 2. Figure 4b indicates the $\alpha$-dependence of the mean DOP, $\rho$, that is obtained from averaging polarization degrees of four to six sheets of CNT micro-cloths prepared at an identical $\alpha$ value.
It is found from Figure 4b that $\rho$ is independent of $\alpha$ even though the $\alpha$-range covers the isotropic-nematic transition line at $\alpha=1$ for denser suspensions of $C_{\mathrm{CNT}}>1$ wt.\% [29,30].
Notably, the constant DOP is ca. 40 \% irrespective of $\alpha=1$.
This is sufficiently high considering that the DOP of super aligned CNT arrays is approximately 50 \% unless the super aligned arrays are further drawn to produce periodically striped films [40-42].
 
\subsection*{Electric field dependencies}

In contrast to Figure 4b, Figure 5a demonstrates that the applied frequency $f$ greatly affects the DOP when the electric field strength $E$ and $\alpha$ are fixed at $E=25$ kV/m and $\alpha=1$, respectively.
In the frequency range of 1 to 20 MHz, we observed a rapid increase in $\rho$ as $f$ is increased, and $\rho$ reached ca. 40 \% at 20 MHz.
The contrasted variations in $\rho$ for Figure 4b and Figure 5a suggest that the anisotropic feature of the CNT micro-cloths is mainly determined by the external AC field, instead of the local alignment degree prior to assembly.
Figure 5b therefore shows frequency dependence in terms of the gathering rates of CNT dispersoids.
Gray scale contrast was used for a time sequence of optical micrographs.
As an index of the gray scale contrast, the longitudinal axis in Figure 5b indicates the normalized mean darkness $G(t^*)$ of the optical micrographs at a predetermined period of time, $t^*=600$ sec, during which the AC field is applied ($E=25$ kV/m and $f=20$ MHz, as before).
Here $G(t)$ is obtained from both spatially averaging the gray scale over the rectangular region of the CNT micro-cloth and normalizing the average to satisfy the condition $G(t)=1$ for black objects. Because $G(t^*)$ represents the total degree of CNT gathering, a smaller value of $G(t^*)$ is associated with a slower assembly rate.

\begin{figure}[hbtp]
\begin{center}
	\includegraphics[
	width=7.5cm
	]{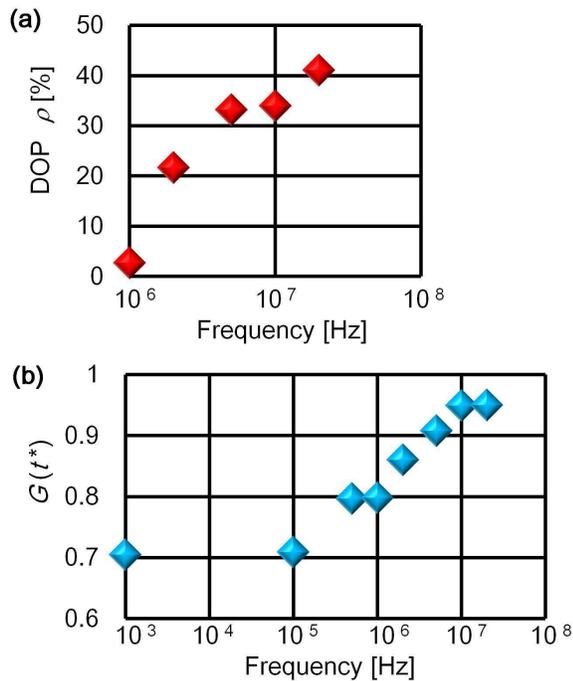}
	\end{center}
	\caption{{\bf Applied frequency ($f$) dependencies.}
We used a DNA-CNT suspension ($\alpha=1$) wile fixing the electric field strength at $E=25$ kV/m.
The $f$-dependences of the DOP $\rho$ ({\bf a}), and the reference darkness $G(t^*)$ at $t^*=600$ sec ({\bf b}) are displayed. Error bars lie within symbols.
}
\end{figure}

Figures 5a,b display a similarity in the upward tendencies of both the DOP and assembly rate:
$\rho$, as well as $G(t^*)$, increases and saturates in an overlapped frequency region.
It is to be noted that previous studies using on-chip electrodes have reported a similar frequency condition ($f>10$ MHz) over which metallic CNTs can be selectively collected [15,16].
The combination of these results implies that the DOP is reduced as the assembly rate is diminished due to the frequency dependence of metallic dielectrophoresis [12-19].
In other words, the decrease in $\rho$ below $f=10$ MHz can be ascribed either to the diminishing polarizability or dielectrophoretic behavior of the metallic CNTs.
However, whether dielectrophoresis is the main cause of CNT collection between the electrode needles remains to be validated.

Hence, we further addressed the assembly mechanism based on a previous formula for the increasing number rate of gathering colloids [39].
This formula predicts that the assembly rate is proportional to $E^2$ when colloids undergo dielectrophoresis toward an electrically focused area.
The $E$-dependence of the assembly rate was investigated using the time evolution of $G(t)$.
In measuring $G(t)$, the present experiments adopted 900 sec as a maximum duration time because it was evaluated from weighing the drops that the open densification for 900 sec yielded a maximal change in the density of the suspension of less than 10 \%.

\begin{figure}[hbtp]
\begin{center}
	\includegraphics[
	width=8.9cm
	]{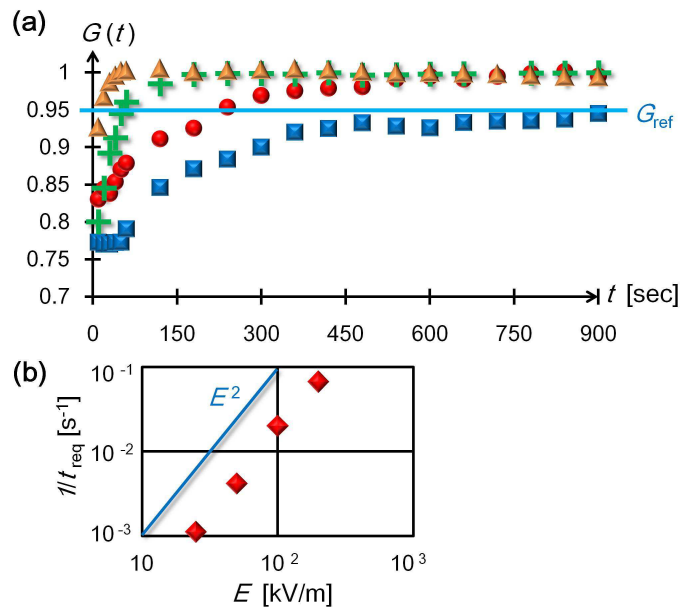}
	\end{center}
	\caption{{\bf Assembly courses dependent on the electric field strength ($E$) in a suspension of DNA-CNT mixture ($\alpha=1$).}
({\bf a}) The normalized mean darkness $G(t)$ of the CNT assembly area is plotted as a function of the duration time, $t$, to compare the time courses for $E=25$ kV/m (blue squares), $E=50$ kV/m (red circles), $E=100$ kV/m (green crosses), and $E=200$ kV/m (orange triangles) at the same frequency of $f=20$ MHz.
({\bf b}) Log-log plot of the inverse required times, $1/t_{\mathrm{req}}$, versus $E$ at $f=20$ MHz.
A light blue line, representing the proportionality of $1/t_{\mathrm{req}}$ to $E^2$, is shown for comparison. Standard errors fall within symbols.
}
\end{figure}

Figure 6a shows four plots of $G(t)$ at field strengths of $E=25,\,50,\,100,\,200$ kV/m and $f=20$ MHz.
Because a maximum value for the normalized gray scale, $G_{\mathrm{ref}}=0.95$, at 900 sec in the lowest electric field strength of $E=25$ kV/m was obtained, we used $G_{\mathrm{ref}}$, which is marked by the shaded blue line in Figure 6a, as an indicator that determines a required time $t_{\mathrm{req}}$ for reaching the threshold of $G_{\mathrm{ref}}$ via the equation, $G(t_{\mathrm{req}})=G_{\mathrm{ref}}$.
From Figure 6a, we can see that the $G(t)$ rises faster with increasing $E$; correspondingly, $t_{\mathrm{req}}$ is reduced by increasing $E$ and the assembly rate can be represented by the inverse time of $1/t_{\mathrm{req}}$.
Figure 6b shows that the experimental data satisfy the proportionality of $1/t_{\mathrm{req}}$ to $E^2$ (the blue line in Figure 6b) consistently with the dielectrophoretic mechanism [17,37,39].

\section*{Conclusions}

In conclusion, we applied a spatially inhomogeneous AC field across DNA-wrapped CNT suspensions by changing the density ratio of $\alpha=C_{\mathrm{CNT}}/C_{\mathrm{DNA}} $ from $\alpha=0.6$ to $\alpha=2$;
incidentally, $\alpha=1$ has been found to be the isotropic-nematic transition line in concentrated suspensions of $C_{\mathrm{CNT}}>1$ wt.\% [33-36].
The inhomogeneous AC fields induce dipole moments in CNTs, especially in metallic tubes, and exerts a dielectrophoretic force, as well as torque, on the CNTs in relatively dilute suspensions (0.1 wt.\%).
This causes the anisotropic CNT-DNA hybrid films to be shaped into 100-$\mu$m scale rectangles, or micro-cloths.
Shaping of the size of these micro-cloths can be accomplished by the electrode needles ({\itshape the on-demand regulation of film size}).
Because this method using micromanipulators is capable of forming assembly films far above the substrate ({\itshape the off-chip assembly}) even at the air-water interface, we can avoid impurity deposition, which is particularly crucial for CNT suspension drying processes that necessarily contain sediments such as depletion-induced aggregates.
Furthermore, thanks to the manipulability of individual electrodes using micromanipulators, the present system allows us to accomplish two operations ({\itshape the one-stop manufacturing}) that other on-chip systems have difficulty performing.
First, we can prepare a dried, free-standing film with one edge attached to an electrode needle, as shown in Figure 2b.
Second, the same film can be subsequently placed over a pinhole, through which a microscope spectrometer can measure the DOP (Movie 3 in Additional File 2).

Finally, we would like to emphasize that, in addition to the maximal achievable DOP value (ca. 40 \%) irrespective of $\alpha$, CNT dispersions containing excess DNA can be a useful medium for maintaining sufficient activity of D-fructose dehydrogenase (FDH) from {\itshape Gluconobacter} sp., an enzyme used in biofuel cells;
indeed, our colorimetric study on oxidation degrees of D-fructose due to FDH \cite{ameyama} shows that the enzymatic activity in pre-assembly suspension of DNA-CNT mixture at $\alpha=2$ dominates that at $\alpha=1$, as well as that in CNT dispersion stabilized by surfactant (Triton X-100) \cite{wagner2}.
Our preliminary result implies that the DNA-stabilized CNT micro-cloths connected directly to an electrode could be useful for biofuel cells, as mentioned previously, due to the direct electron transfer and/or improved enzymatic activity.

\section*{Movie captions}

{\bf Movie 1}:
A typical sequence of preparing anisotropic DNA-CNT micro-cloth that consists of three steps: an electric gathering of aligned carbon nanotubes between a parallel pair of electrode needles, vaporization when the air-water interface moves across the assembly surface, and separation of one needle from the micro-cloth while the other electrode remains fixed to one edge of the assembly film.

{\bf Movie 2}:
A vaporization-induced flow washing out two rectangular edges, opposite to each other, more extensively than those in Movie 1. Upon evaporation, the CNT assembly is separated from one electrode and is touched by the free needle itself, with the other electrode remaining adherent to the CNT micro-cloth.

{\bf Movie 3}:
In the early part of this movie, the location of a dried CNT film is adjusted by manipulating an attached electrode needle so that the black sheet may cover a pinhole which can be seen as a bright circle. Subsequently, the film brightness varies according to the rotation of a polarizing plate.

\subsection*{Acknowledgements}

This work was funded by the Yamada foundation.

\end{document}